# NEW EQUATIONS FOR SEA WATER DENSITY CALCULATION BASED ON MEASUREMENTS OF THE SOUND SPEED

A N Grekov[1], N A Grekov[2], E Sychov[3]

[1] *Institute of Natural and Technical Systems, Laboratory of field monitoring systems, 28 Lenina street, 299011 Sevastopol, Russia, grekov@protonmail.com.*
[2] *Institute of Natural and Technical Systems, Special scientific design and technology bureau, 28 Lenina street, 299011 Sevastopol, Russia, ngrekov1@ya.ru.*
[3] *Institute of Natural and Technical Systems, Laboratory of field monitoring systems, 28 Lenina street, 299011 Sevastopol, Russia, sychov-e@rambler.ru.*

**Abstract:**

Density is one of the most important properties of seawater and is used in various marine research and technology. Traditionally, in the practice of oceanographic research, it is customary to consider density as a dependent parameter, which is a function of several other parameters taken as independent. Usually the following three parameters are used as the independent parameters: temperature, hydrostatic pressure and salinity. The issues of temperature and hydrostatic pressure measuring in situ are technologically well developed, while in the salinity measuring there are still unsolved problems. This is due to the fact that salinity is such a property that it is simply impossible to determine directly in situ. To eliminate the problems associated with measurements of salinity, the authors developed the special new kind equation. That equation of the new kind express the density of sea water through independent and in situ measured parameters: temperature, hydrostatic pressure and sound speed. The novelty of this approach is that using of the sound speed as the independent parameter makes it possible to exclude measurements of salinity. The authors developed two such new equations for the different cases of using. The first new equation is intended for use in technical applications and reproduces the sea water density in a wide range of the aquatic environment parameters with a root mean square deviation of 0.062 kg/m$^3$. The second more precise new equation is intended for scientific applications and reproduces the sea water density in a narrower oceanographic range of parameters with a root mean square deviation of 0.0018 kg/m$^3$.

**Keywords:** equation of state, sea water, density, sound velocity, TEOS-10, root mean square deviation.

## Introduction.

The sea water density is one of its most important properties and is used in the calculation of geostrophic currents, modeling of the sea water movement dynamics in various marine technologies and applications.

It is known that the sea water density is equal to the mass of the sea water unit volume, which is fixed under certain conditions, i.e. under condition of constancy of a number of independent parameters. In the practice of oceanographic studies the following three parameters are normally used: the temperature, pressure (hydrostatic) and salinity [1]. The issues of temperature and pressure measuring in situ are technologically well developed, while in the salinity measuring there are still unsolved problems. This is due to the fact that salinity is such a property that it is simply impossible to determine directly in situ. The traditional laboratory chemical analysis of sea water with sampling from various depths is a long, time-consuming, complex, and costly process. Therefore, salinity is usually determined indirectly through the results of measuring of any physical index of seawater, for example, based on the electric conductivity of the sea water or the sea water refractive index [2]. Between salinity and conductivity a functional relationship is established with a reasonable accuracy. Currently, the following technology for absolute salinity determination has been adopted in the practice of oceanographic research: first, based on the conductivity ratio, whose measurement technique has not yet been properly metrologically assured [3,4,5], the relative salinity value is determined, which is then corrected by adding a special correction to the absolute salinity value [6]. Thus, a somewhat insufficiently known correction is added to the not completely justified salinity value. Specifically, for this correction, based on the results of laboratory measurements of the sea water density in 811 samples taken in the Atlantic (1 point, 9 samples), in the Pacific (1 point, latitudinal section – 22 points, two meridional sections 11 and 19 points each, 585 samples) and in the Indian (1 point, one meridional section – 48 points, 217 samples) Oceans, and also using approximate relationships between anomalies of absolute salinity and silicate concentrations, the global correction atlas for absolute salinity was calculated [7]. It is obvious that the validity of such an atlas leaves much to be desired. The existence of problems related to the accuracy of the salinity measurement is also recognized by the developers of the international seawater state equation TEOS-10 [4,5]. The approach they proposed to the problem solution is apparently due to the desire to make only a correction to the results of salinity indirect measurements without changing the role and significance of salinity measurements as a whole. Thus, in [8], the authors checked the long-term constancy of the salt composition of the standard sea water in the North Atlantic, a new density-salinity relation based on accurate density measurements was presented and an assumption was made that the density accuracy in TEOS-10 was overestimated. In [9], the authors proposed a new equation, consisting of 69 terms, for the calculation of salinity based on measurements of sound velocity, temperature, and pressure. The authors of all the works opinion is that the role and significance of salinity indirect measurements which are now practically of a monopoly nature, will have to decrease as more accurate methods of in situ measuring of other parameters of the marine environment develop.

## Theoretical justification.

From the point of view of the solutions thermodynamics, sea water is a multicomponent solution formed by water molecules and substances dissolved in it, consisting of dozens (more than 40) of chemical elements. However, in reference books on the sea water salt composition, only 15 main components are taken into account, mainly in the form of ions (cations and anions), whose mass fraction exceeds 0.001 g/kg [10], and the influence of the remaining components is generally ignored. Therefore, taking into account the above, it can be assumed that seawater is a 16-component (15+1) system.

According to the Gibbs phase rule, the number of independent intensive (independent of the matter quantity) variables (state parameters) that determine the thermodynamic system equilibrium, is equal to

$p = n – m + 2,$ (1)

where n – the number of the system components; m – the number of phases in the system; 2 – number of freedom degrees.

In our case, for the sea water, we obtain that p = 16 – 1 + 2 = 17. This means that any sea water intensive parameter is uniquely determined by the value of 17 other intensive parameters, i.e.

$Y_1 = f_{Y1}(Y_2, Y_3, x_1, \dots, x_{15}),$ (2)

where $Y_i$ – three of any intense thermodynamic parameters; $x_j$ – the mass fractions of 15 main components of the sea water salt part. Equation (2) is called the equation of state of the given system (in our case, the sea water). It can also be represented in the form of a homogeneous equation

$F(Y_1, Y_2, Y_3, x_1, \dots, x_{15}) = 0.$ (3)

By way of example, let us write the equations of state for density (ρ) and sound velocity (c) depending on temperature, pressure (hydrostatic), and mass fractions of the sea water main components:

$F_\rho(\rho, T, P, x_1, \dots, x_{15}) = 0,$ (4)

$F_c(c, T, P, x_1, \dots, x_{15}) = 0.$ (5)

Obviously, the equations of state in the form (4) or (5) cannot be realized in practice because of the huge number of independent parameters. However, the way out was found by replacing 15 parameters expressing the mass fractions of the 15 main components of the sea water salt part with only one parameter known as the sea water salinity.

$S = S(x_1, \dots, x_{15}).$ (6)

The justification for such a replacement is the law discovered at the end of the XIX century during the famous scientific expedition of the Challenger vessel, the law is known as 'constancy of the salt composition' (more precisely, the constancy of the structure of the salt (ionic) composition), also called by the author's name 'Dittmar`s law ', according to which: 'The quantitative ratio of the main ions in the sea water always remains unchanged in different regions of the oceans, and the share of the remaining substances is so small that it may not be taken into account when carrying out various hydrochemical studies' [11]. This discovery was made based on the result of analysis of a very small number of samples (total 77 samples) of the sea water taken from different regions of the World Ocean. We should note from the very beginning that this law contains a number of assumptions and is therefore not strict; moreover, recent studies have revealed the presence of non-typical anomalous zones in the seas and oceans, where this law is either violated or not observed at all. Suffice it to say, that the waters of such seas as the Black Sea, the Sea of Azov, the Baltic, the Caspian, the Aral Sea differ in their ionic composition both among themselves and the World Ocean [3,12]. Therefore, we can state that the above-mentioned law of constancy of the ionic composition of the sea water is not of universal, but of local nature limited by bays, seas and other geograph-ically limited bodies of water that manage to come either to a state of thermodynamic equilibrium or to a state close to that.

Taking into account (6) equations (4) and (5) we can put it down in the following form:

$F_\rho(\rho, T, P, S) = 0,$ (7)

$F_c(c, T, P, S) = 0.$ (8)

In our works [13,14], the form (8) was successfully used to construct the equation of the sound velocity. Equations (7) and (8) obey Dittmar`s law. From equation (8), one can obtain an equation expressing the salinity

$S = f_S(c, T, P).$ (9)

Substituting (9) into equation (7), we obtain:

$F_\rho(\rho, T, P, f_S(c, T, P)) = 0$ (10)

or

$F_\rho(\rho, T, P, c) = 0.$ (11)

We should note, that the values of the density and sound velocity are parametrically related here to the structure of salinity. Thus, the equations for the density in the form (10) or (11), although, not containing the salinity in explicit form, nevertheless, also obey Dittmar`s law.

Equations for the density in the form (10) or (11) can be very convenient for use in the practice of marine research, since the presence of such equations makes it possible to replace the measuring channel of electrical conductivity with a measuring channel of sound velocity, which, from the metrological point of view is more justified [15]. The use of equations (10) or (11), in which the input parameters are the data of the measuring channels of sound velocity, temperature and pressure, will allow us to quickly determine the density of the sea water with high accuracy without measuring the salinity.

The fulfillment of equation (6) is equivalent to the fact that sea water is no longer regarded as a multicomponent system, but as a two-component system for which the number of independent intense state parameters would be equal to $p = 2 – 1 + 2 = 3$. Therefore, we can write the equation of the sea water state for any combination of only four (3+1) intensive parameters, which even do not include salinity. As a result, for the parameters T, P, c, and ρ we immediately arrive at the following form of the equation of state in a homogeneous form

$\Phi(T, P, c, \rho) = 0,$ (12)

where for the sound density and velocity expressed explicitly, we can put down, respectively:

$\rho = \varphi_\rho(T, P, c),$ (13)

$c = \varphi_c(T, P, \rho).$ (14)

There is a fundamental difference between the equations of state (11) and (12), despite their external similarity. The equation of state (12) and its derivative forms (13) and (14) differ in that for them it is not actually necessary to impose such a serious restriction as the coincidence of the sea water ionic composition structure with the international standard [4]. Therefore, unlike the equations of state in the form (7) and (8) or (10) and (11), they can find application also in anomalous or non-typical zones where Dittmar`s law does not work. To establish the form of the functional dependence in equations (12) or (13) and (14), it is sufficient to have the results of synchronous in situ measurements of temperature, pressure, sound velocity and density. If in situ measurement issues for first three of the listed parameters are currently technologically developed, then the question of in situ measuring the sea water density is still in the development stage.

In the event, attempts to develop a methodology for in situ measuring the sea water density are successful, and the prerequisites for this are already available, the emphasis in the practice of marine measurements can gradually shift from salinity measurements toward measurements of density and sound velocity performed both synchronously and independently.

**Equation I of technical purposes sea water density (in a wide range of parameters).**

To generate the equation of state of the form (13), the authors used the international TEOS-10 system [4]. Based on the calculations for the equations for the density and sound velocity of the TEOS-10 system [4], an array of initial data of the form $\{\rho_m, c_m, T_m, P_m, S_m\}_M$ was generated in which 'M', the number of points, totaled more than 130 thousand. To this end, the values of 'M' pairs of densities ρ ($T_m$, $P_m$, $S_m$) and sound velocities c ($T_m$, $P_m$, $S_m$) were calculated with step of 1°C in the temperature range from the melting curve to 40°C with a step of 2 MPa in the range hydrostatic pressures 0 - 120 MPa and with a step of 1 g/kg in the salinity range 0 – 42 g/kg. At the same time, the range of sound velocity variation made about 1,300 - 1,800 m/s, and the density 990 – 1,090 kg/m$^3$, respectively. In a set of five parameters (ρ, c, T, P, S), only three, in any combination, can be used as independent.

The proposed equation for interpolating the sea water density of type (13) was adopted in the following form

$$\gamma = \sum_i \sum_j \sum_k b_{ijk} \tau^i \pi^j \omega^k \qquad (15)$$

where $\gamma = (\rho - \rho_0)/\rho^*$; $\tau = (T - T_0)/T^*$; $\pi = (P_{абс} - P_0)/P^*$; $\omega = (c - c_0)/c^*$; $\rho_0$ = 990 kg/m$^3$; $\rho^*$ = 100 kg/m$^3$; $T_0$ = -10°C; $T^*$ = 50°C; $P_0$ = 0,101325 MPa; $P^*$ = 120 MPa; $c_0$ = 1300 m/s; $c^*$ = 500 m/s.

Equation (15) expresses the functional dependence of the sea water density on the parameters T, P, and c in explicit form and can be useful in marine research. The advantage of the equation is that salinity is not clearly present in the equation, although it is implied that such properties of the sea water as density and sound velocity are functionally related to salinity. The presence of an equation of this kind makes it possible to exclude indirect (via conductivity) measurements of salinity, replacing them by direct measurements of the sound velocity in a number of cases.

For equation I of seawater density, an index matrix is used in the form (15) which contains 79 non-zero coefficients. The index matrix for equation I of sea water density is given in Table. 1. 1.

**Table 1.** Index matrix of the sea water density equation I in the form (15).

| k = | 0 | 0 | 0 | 0 | 0 | 1 | 1 | 1 | 1 | 1 | 2 | 2 | 2 | 2 | 3 | 3 | 3 | 4 | 4 | 4 |
|---|---|---|---|---|---|---|---|---|---|---|---|---|---|---|---|---|---|---|---|---|
| j = | 0 | 1 | 2 | 3 | 4 | 0 | 1 | 2 | 3 | 4 | 0 | 1 | 2 | 3 | 0 | 1 | 2 | 0 | 1 | 2 |
| i = 0 | + | + | + | + | + | + | + | + | + | + | + | + | + | + | + | + | + | + | + | + |
| i = 1 | + | + | + | + | + | + | + | + | + | + | + | + | + | + | + | + | + | + | + | + |
| i = 2 | + | + | + | + | + | + | + | + | + | + | + | + | + | + | – | – | – | – | – | – |
| i = 3 | + | + | + | + | + | + | + | + | + | + | – | – | – | – | – | – | – | – | – | – |
| i = 4 | + | + | + | + | + | + | + | + | + | + | – | – | – | – | – | – | – | – | – | – |
| i = 5 | + | + | + | + | + | – | – | – | – | – | – | – | – | – | – | – | – | – | – | – |

In this index matrix, on each lines and columns' intersection, only one nonrecurrent combination of indices can be obtained. Here the sign '+' means that the given combination of indices is used, and the sign '–' means that this combination of indices is omitted. In total, 79 combinations were used in this index matrix, i.e., equation (15) in this case will contain 79 coefficients.

The coefficients $b_{ijk}$ of the density equation I in the form (15) were calculated by the authors using the least squares method and together with the indices are given in table. 2.

**Table 2.** Indices and coefficients of the equation I for the sea water density.

| N | i | j | k | $b_{ijk}$ | N | i | j | k | $b_{ijk}$ |
|---|---|---|---|---|---|---|---|---|---|
| 1 | 0 | 0 | 0 | -0,323474572353243 | 41 | 0 | 2 | 1 | 41,1122768956598 |
| 2 | 1 | 0 | 0 | -2,61052852853926 | 42 | 1 | 2 | 1 | 127,489244347504 |
| 3 | 2 | 0 | 0 | -1,25282282933016 | 43 | 2 | 2 | 1 | 90,4770873310821 |
| 4 | 3 | 0 | 0 | -4,05194768329123 | 44 | 3 | 2 | 1 | 34,0568943202623 |
| 5 | 4 | 0 | 0 | -4,75718928994638 | 45 | 4 | 2 | 1 | -2,80647766031511 |
| 6 | 5 | 0 | 0 | 1,35246306700429 | 46 | 0 | 3 | 1 | 26,0700206023845 |
| 7 | 0 | 1 | 0 | -1,58573785665124 | 47 | 1 | 3 | 1 | 26,8962331714441 |
| 8 | 1 | 1 | 0 | -6,17802678326577 | 48 | 2 | 3 | 1 | 64,8467249538919 |
| 9 | 2 | 1 | 0 | -17,9414321072584 | 49 | 3 | 3 | 1 | -6,23805014385411 |
| 10 | 3 | 1 | 0 | -29,5016465982731 | 50 | 4 | 3 | 1 | 5,78653868844489 |
| 11 | 4 | 1 | 0 | 10,0361583156344 | 51 | 0 | 4 | 1 | -14,7238013318731 |
| 12 | 5 | 1 | 0 | -1,02367871521456 | 52 | 1 | 4 | 1 | 15,5233415705374 |
| 13 | 0 | 2 | 0 | -3,94271532912074 | 53 | 2 | 4 | 1 | -5,00221702113707 |
| 14 | 1 | 2 | 0 | -17,3180226870909 | 54 | 3 | 4 | 1 | -2,38569664017924 |
| 15 | 2 | 2 | 0 | -56,1941458233329 | 55 | 4 | 4 | 1 | 1,30899162542083 |
| 16 | 3 | 2 | 0 | -10,2232404603107 | 56 | 0 | 0 | 2 | -26,0896213783704 |
| 17 | 4 | 2 | 0 | 1,08008089604956 | 57 | 1 | 0 | 2 | -81,7959395776633 |
| 18 | 5 | 2 | 0 | -0,215204425138987 | 58 | 2 | 0 | 2 | -109,219751364335 |
| 19 | 0 | 3 | 0 | -5,91559663215341 | 59 | 0 | 1 | 2 | -107,08543599038 |
| 20 | 1 | 3 | 0 | -29,2676693238212 | 60 | 1 | 1 | 2 | -227,106998190509 |
| 21 | 2 | 3 | 0 | -33,8072052796647 | 61 | 2 | 1 | 2 | -2,47721750331145 |
| 22 | 3 | 3 | 0 | 0,0485154669118931 | 62 | 0 | 2 | 2 | -85,8422506476092 |
| 23 | 4 | 3 | 0 | -4,76287215989976 | 63 | 1 | 2 | 2 | 93,3545959547231 |
| 24 | 5 | 3 | 0 | 0,478522498835545 | 64 | 2 | 2 | 2 | -131,655865531367 |
| 25 | 0 | 4 | 0 | -3,70090732710256 | 65 | 0 | 3 | 2 | 92,3924544521505 |
| 26 | 1 | 4 | 0 | -11,8242581200948 | 66 | 1 | 3 | 2 | -48,2580115742852 |
| 27 | 2 | 4 | 0 | -4,00674681588644 | 67 | 2 | 3 | 2 | 16,0335487887494 |
| 28 | 3 | 4 | 0 | -0,0305307923202697 | 68 | 0 | 0 | 3 | 97,3052961781545 |
| 29 | 4 | 4 | 0 | -1,04873066115089 | 69 | 1 | 0 | 3 | 179,263777041781 |
| 30 | 5 | 4 | 0 | -0,81806496328397 | 70 | 0 | 1 | 3 | 173,550444094841 |
| 31 | 0 | 0 | 1 | 6,05198875817299 | 71 | 1 | 1 | 3 | -193,929771714651 |
| 32 | 1 | 0 | 1 | 14,3793233043429 | 72 | 0 | 2 | 3 | -201,925242746972 |
| 33 | 2 | 0 | 1 | 25,7412632398216 | 73 | 1 | 2 | 3 | -20,6900315526571 |
| 34 | 3 | 0 | 1 | 37,9676817625738 | 74 | 0 | 0 | 4 | -142,091866687502 |
| 35 | 4 | 0 | 1 | -4,52965127701382 | 75 | 1 | 0 | 4 | 9,68482361465972 |
| 36 | 0 | 1 | 1 | 19,1925176879528 | 76 | 0 | 1 | 4 | 141,954590076825 |
| 37 | 1 | 1 | 1 | 73,4509857972438 | 77 | 1 | 1 | 4 | 95,6897154048258 |
| 38 | 2 | 1 | 1 | 152,438792774745 | 78 | 0 | 2 | 4 | 3,08061726293501 |
| 39 | 3 | 1 | 1 | -8,22361002943686 | 79 | 1 | 2 | 4 | 0,402558073773966 |
| 40 | 4 | 1 | 1 | -2,7406780332077 | – | – | – | – | – |

The root mean square deviation (RMSD) over the entire array of initial data (more than 130 thousand points) for the density equation I equals to 0.062 kg/m$^3$. The calculated deviations histogram given in table 3, confirms the normal law of deviations distribution. The number of computational points in which deviations exceed 3$\sigma$, makes about 1%.

**Table 3.** Histogram of the design density deviations for the sea water density equation I.

| 1 | 2 | 3 | 4 | 5 | 6 | 7 | 8 | 9 | 10 | |
|---|---|---|---|---|---|---|---|---|---|---|
| <-0,24 | <-0,18 | <-0,12 | <-0,06 | <0 | <0,06 | <0,12 | <0,18 | <0,24 | <+∞ | Total: |
| 0,24444 | 0,58681 | 1,8225 | 5,9459 | 43,35 | 40,125 | 5,6544 | 1,0936 | 0,46571 | 0,71165 | 100% |

A graphic representation of the sea water design density deviations histogram over the entire array of initial data of density equation I is shown in fig. 1.

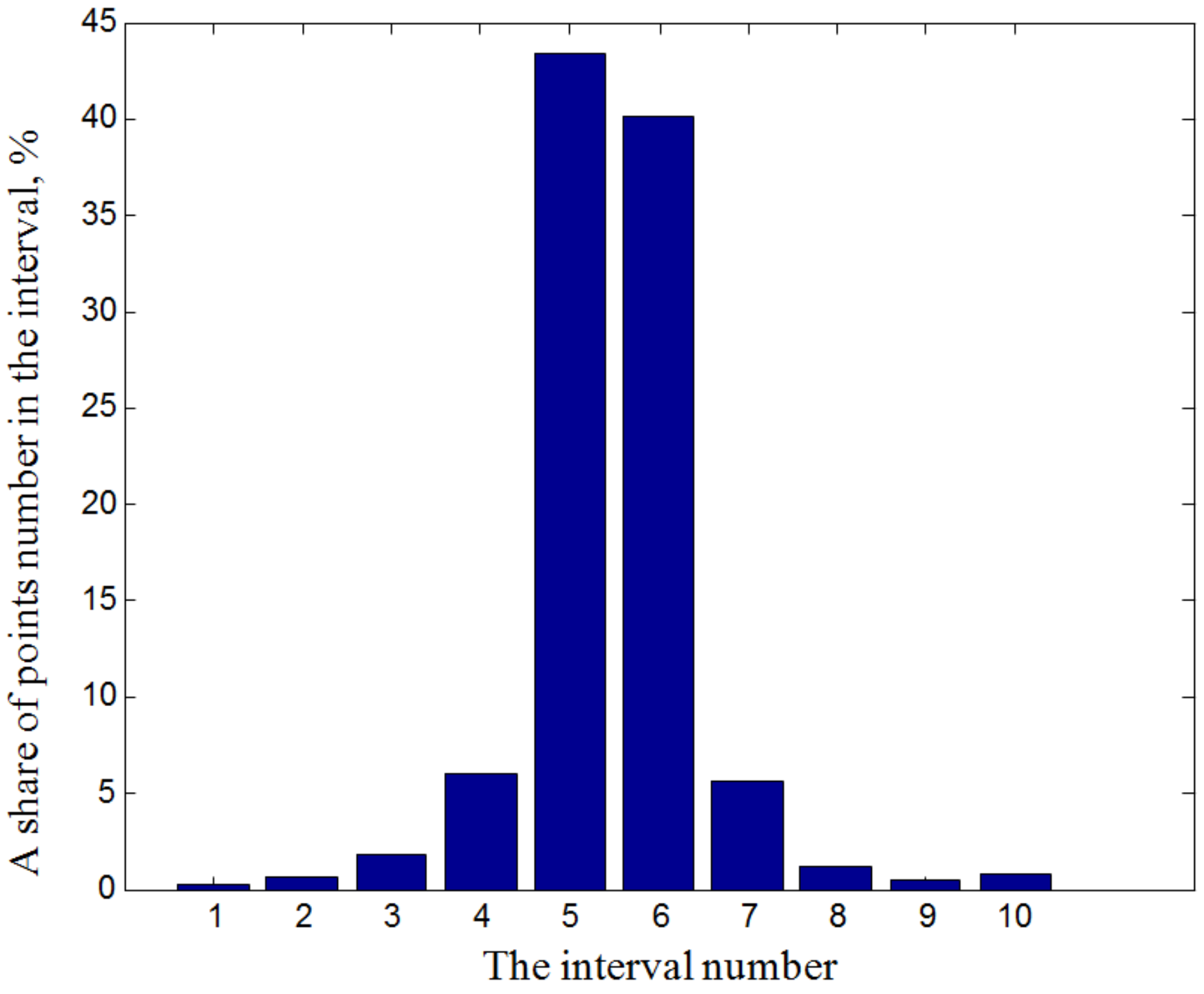

**Figure 1:** histogram of deviations for the density equation I

The range of parameters for which the density equation I is constructed is wider than the oceanographic one, and therefore, can be used for waters with anomalous properties, for example, located in the region of hydrothermal sources outputs. This equation is also suitable for calculating the density of fresh and slightly saline waters at temperatures up to 40°C.

**Density equation II for scientific purposes (in the oceanographic range of parameters).**

The density equation II is developed for a narrower and more probable from the point of view of the practical implementation of the oceanographic (' Neptunian ') range of sea waters parameters. The boundary values of temperature and salinity in the oceanographic parameters range depending on hydrostatic pressure are shown in fig. 2. Temperature $T_{min} = min(T_{melt}, -2^oC)$ was accepted as the minimal.

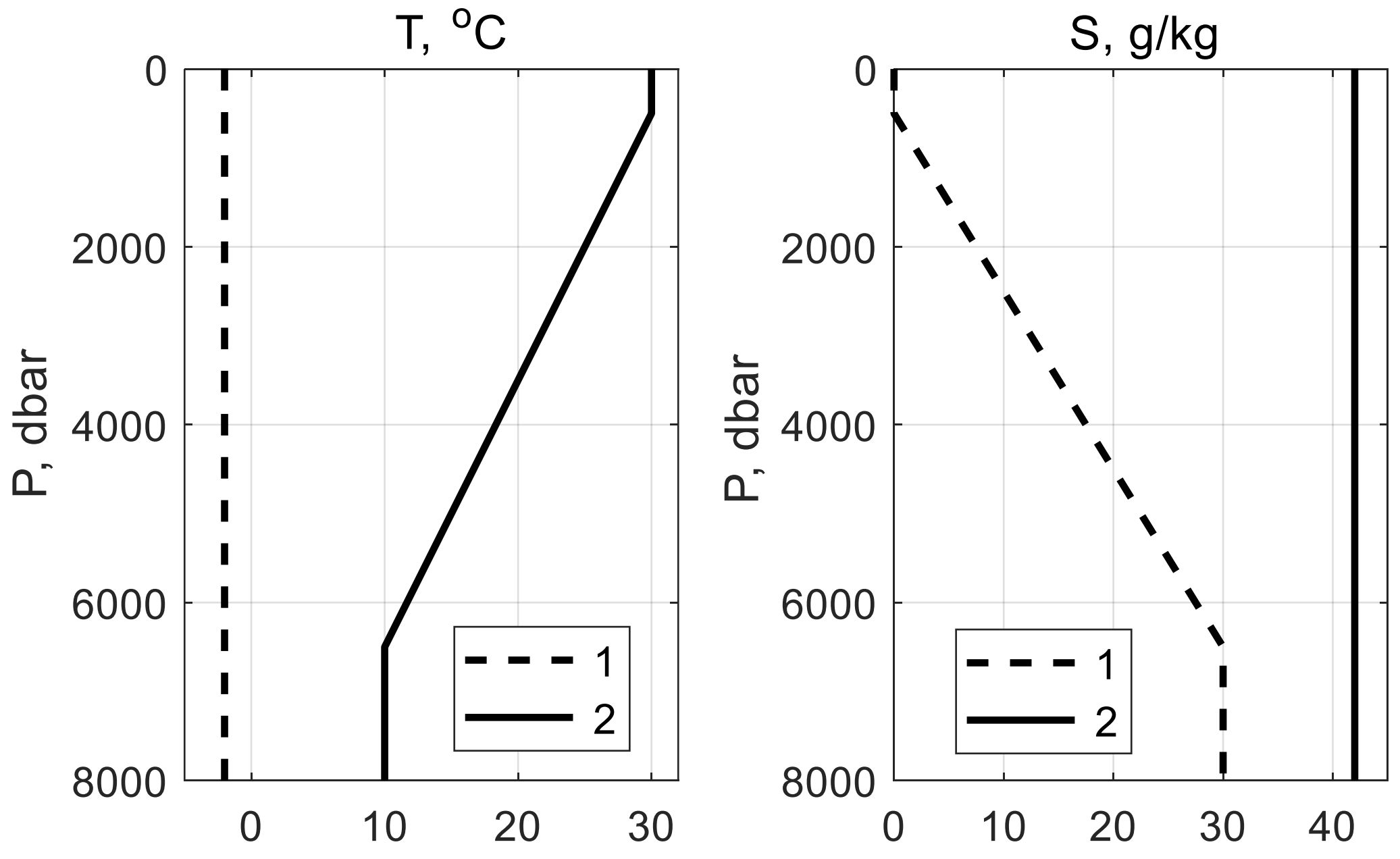

a) b)

**Figure 2:** the boundary values of temperature (a) and salinity (b) in the oceanographic parameters range depending on the hydrostatic pressure: 1 and 2 - the boundaries of the minimal and maximal parameter values.

To construct the equation II of sea water density, the authors used an array of initial data based on density, temperature, hydrostatic pressure, and sound velocity in an amount of about 200,000 points. The data belong to the oceanographic parameters range and are located at the nodes of a sufficiently dense three-dimensional grid of independent parameters: in the pressure range from 0 to 80 MPa with a step of 0.25 MPa, in the temperature range from $T_{min}$ to $T_{max}(P_M)$ with a step of 1°C, in the salinities range from $S_{min}(P_M)$ to 42 g/kg with a step of 1 g/kg. To prepare the initial data array, the international TEOS-10 system was used [4].

The equation II of sea water density developed by the authors was adopted in the above-described polynomial form (15). Equation II uses its own index matrix containing 80 non-zero coefficients. The index matrix for equation II of sea water density is given in table. 4.

**Table 4.** Index matrix of the equation II of the sea water density in the form (15).

| k = | 0 | 0 | 0 | 0 | 0 | 1 | 1 | 1 | 1 | 1 | 2 | 2 | 2 | 2 | 2 | 3 | 3 | 3 | 4 | 4 | 4 | 5 | 5 | 6 |
|---|---|---|---|---|---|---|---|---|---|---|---|---|---|---|---|---|---|---|---|---|---|---|---|---|
| j = | 0 | 1 | 2 | 3 | 4 | 0 | 1 | 2 | 3 | 4 | 0 | 1 | 2 | 3 | 4 | 0 | 1 | 2 | 0 | 1 | 2 | 0 | 1 | 0 |
| i = 0 | + | + | + | + | + | + | + | + | + | + | + | + | + | + | + | + | + | + | + | + | + | + | + | + |
| i = 1 | + | + | + | + | – | + | + | + | + | – | + | + | + | + | – | + | + | – | + | + | – | + | + | – |
| i = 2 | + | + | + | + | – | + | + | + | – | – | + | + | + | – | – | + | + | – | + | – | – | + | – | – |
| i = 3 | + | + | + | – | – | + | + | + | – | – | + | + | – | – | – | + | – | – | + | – | – | – | – | – |
| i = 4 | + | + | – | – | – | + | + | – | – | – | + | + | – | – | – | + | – | – | – | – | – | – | – | – |
| i = 5 | + | – | – | – | – | + | + | – | – | – | + | – | – | – | – | – | – | – | – | – | – | – | – | – |
| i = 6 | + | – | – | – | – | + | – | – | – | – | + | – | – | – | – | – | – | – | – | – | – | – | – | – |

The coefficients $b_{ijk}$ of the density equation II in the form (15) were calculated by the authors using the least squares method and these coefficients are given in table 2. together with the indices.

**Table 5.** Indices and coefficients of the sea water density equation II

| N | i | j | k | $b_{ijk}$ | N | i | j | k | $b_{ijk}$ |
|---|---|---|---|---|---|---|---|---|---|
| 1 | 0 | 0 | 0 | -0.342846942821053 | 41 | 0 | 0 | 2 | -41.6554540377067 |
| 2 | 1 | 0 | 0 | -2.37298311124372 | 42 | 1 | 0 | 2 | -47.034820188248 |
| 3 | 2 | 0 | 0 | 5.81288509909762 | 43 | 2 | 0 | 2 | 170.91454147007 |
| 4 | 3 | 0 | 0 | 7.6797860277669 | 44 | 3 | 0 | 2 | -4.82693133113655 |
| 5 | 4 | 0 | 0 | -10.2712686927078 | 45 | 4 | 0 | 2 | 96.6666431835976 |
| 6 | 5 | 0 | 0 | 15.8107586603892 | 46 | 5 | 0 | 2 | 7.25415143121054 |
| 7 | 6 | 0 | 0 | -3.30901780030299 | 47 | 6 | 0 | 2 | -4.1316752598705 |
| 8 | 0 | 1 | 0 | -1.94945253994349 | 48 | 0 | 1 | 2 | -229.318151417977 |
| 9 | 1 | 1 | 0 | -3.47487625741016 | 49 | 1 | 1 | 2 | -277.228152351038 |
| 10 | 2 | 1 | 0 | 14.8675430180194 | 50 | 2 | 1 | 2 | 127.465803646146 |
| 11 | 3 | 1 | 0 | -22.1676882261462 | 51 | 3 | 1 | 2 | 46.6401758107173 |
| 12 | 4 | 1 | 0 | 26.5328176360615 | 52 | 4 | 1 | 2 | 40.2441229325806 |
| 13 | 0 | 2 | 0 | -4.51819721676465 | 53 | 0 | 2 | 2 | -240.8234267418 |
| 14 | 1 | 2 | 0 | 5.74357544141893 | 54 | 1 | 2 | 2 | 160.07280714177 |
| 15 | 2 | 2 | 0 | -2.48388404592973 | 55 | 2 | 2 | 2 | 77.5908225662814 |
| 16 | 3 | 2 | 0 | 14.6230575549256 | 56 | 0 | 3 | 2 | 99.4081607406654 |
| 17 | 0 | 3 | 0 | 1.56264563441416 | 57 | 1 | 3 | 2 | 76.1815189240049 |
| 18 | 1 | 3 | 0 | 11.8795971115739 | 58 | 0 | 4 | 2 | 27.1375934310002 |
| 19 | 2 | 3 | 0 | 4.36938691162792 | 59 | 0 | 0 | 3 | 229.325929866316 |
| 20 | 0 | 4 | 0 | 7.36733982806682 | 60 | 1 | 0 | 3 | 292.12247103212 |
| 21 | 0 | 0 | 1 | 6.9939003370351 | 61 | 2 | 0 | 3 | -162.697052110626 |
| 22 | 1 | 0 | 1 | 7.21903046174993 | 62 | 3 | 0 | 3 | -21.244143156862 |
| 23 | 2 | 0 | 1 | -63.7519001289791 | 63 | 4 | 0 | 3 | -25.0052738596446 |
| 24 | 3 | 0 | 1 | -2.5048565449611 | 64 | 0 | 1 | 3 | 888.225752060885 |
| 25 | 4 | 0 | 1 | -14.8904378729029 | 65 | 1 | 1 | 3 | 712.367745016693 |
| 26 | 5 | 0 | 1 | -41.8310125895913 | 66 | 2 | 1 | 3 | -51.3125481496336 |
| 27 | 6 | 0 | 1 | 10.7954513447102 | 67 | 0 | 2 | 3 | 505.196806480832 |
| 28 | 0 | 1 | 1 | 29.6670561099301 | 68 | 0 | 0 | 4 | -734.238729145698 |
| 29 | 1 | 1 | 1 | 38.4930782208892 | 69 | 1 | 0 | 4 | -774.648010499607 |
| 30 | 2 | 1 | 1 | -37.2422478531876 | 70 | 2 | 0 | 4 | 33.3926996176371 |
| 31 | 3 | 1 | 1 | -0.251713713665298 | 71 | 3 | 0 | 4 | -40.859663102508 |
| 32 | 4 | 1 | 1 | -74.6902894673729 | 72 | 0 | 1 | 4 | -1702.82855445625 |
| 33 | 5 | 1 | 1 | 3.29863591975545 | 73 | 1 | 1 | 4 | -868.841595240282 |
| 34 | 0 | 2 | 1 | 51.2707717241145 | 74 | 0 | 2 | 4 | -593.611624424474 |
| 35 | 1 | 2 | 1 | -25.4114531734917 | 75 | 0 | 0 | 5 | 1248.67338861614 |
| 36 | 2 | 2 | 1 | -86.8100171882373 | 76 | 1 | 0 | 5 | 731.757936454975 |
| 37 | 3 | 2 | 1 | -16.6837599120359 | 77 | 2 | 0 | 5 | -19.5665209604199 |
| 38 | 0 | 3 | 1 | -23.6449621178942 | 78 | 0 | 1 | 5 | 1398.54321562531 |
| 39 | 1 | 3 | 1 | -99.1956573575483 | 79 | 1 | 1 | 5 | 63.6119012778081 |
| 40 | 0 | 4 | 1 | -39.6940104014377 | 80 | 0 | 0 | 6 | -877.036789516273 |

The root mean square deviation (RMSD) σ for the entire initial data array (about 200 thousand points) for the density equation II is 0.0018 kg/m$^3$, which is more than twice lower than the RMSD of laboratory data (0.004 kg/m$^3$) used in the international system TEOS-10 [4]. The calculated deviations histogram given in table 6, confirms the normal law of deviations distribution. The number of computational points in which deviations exceed 3σ, makes about 1%.

**Table 6.** Histogram of the design density deviations for the equation II of the sea water density.

| 1 | 2 | 3 | 4 | 5 | 6 | 7 | 8 | 9 | 10 | |
|---|---|---|---|---|---|---|---|---|---|---|
| <-0.008 | <-0.006 | <-0.004 | <-0.002 | <0 | <0.002 | <0.004 | <0.006 | <0.008 | <+∞ | Total: |
| 0.2902 | 0.1641 | 0.9061 | 7.3787 | 40.4739 | 42.5567 | 6.2215 | 1.3304 | 0.4838 | 0.1946 | 100% |

A graphical representation of the histogram of the sea water design density deviations over the entire array of initial data of the density equation II is shown in fig. 3.

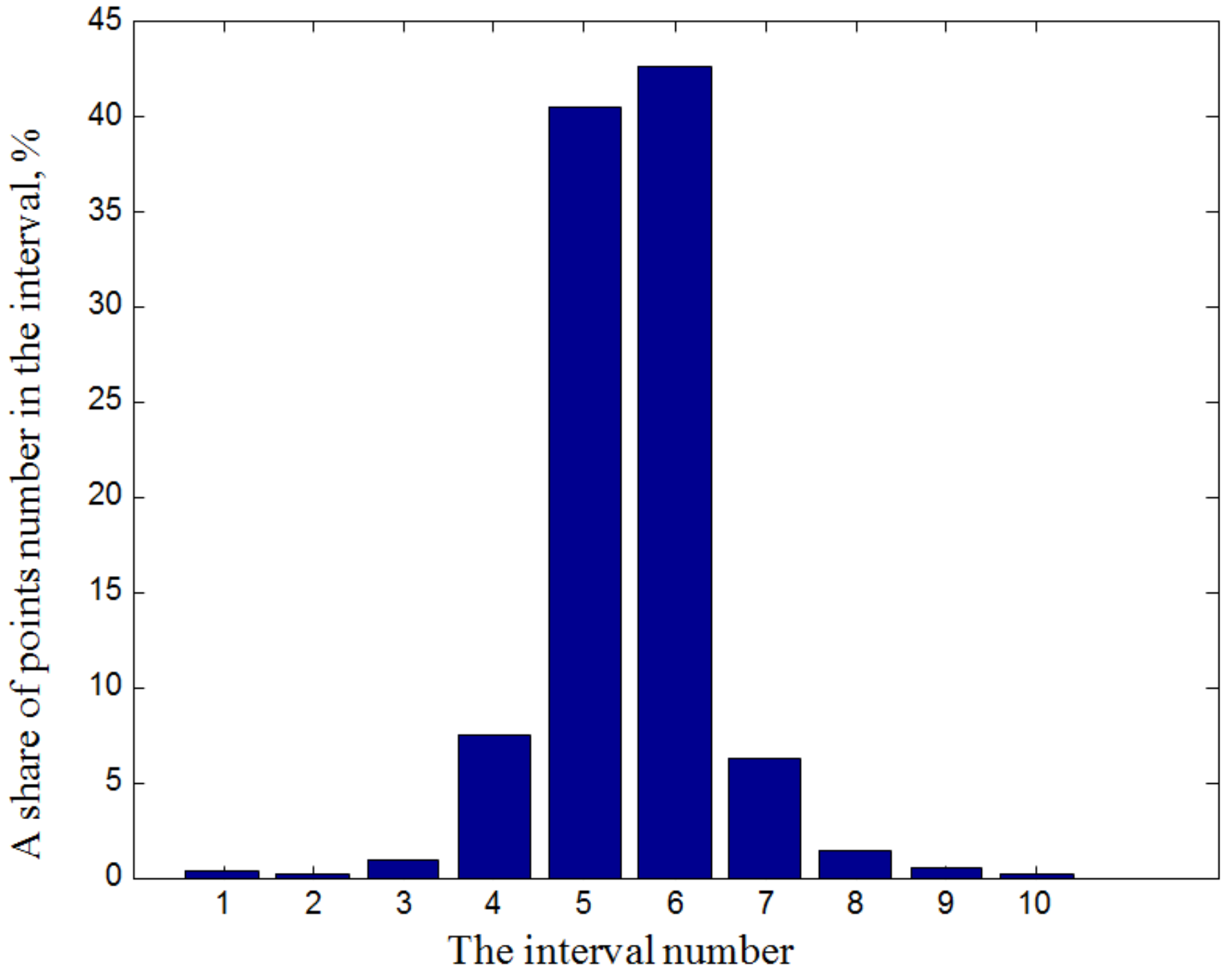

**Figure 3:** histogram of deviations for equation II.

**Results analysis.**

The new equation of state of new type (13) obtained by the authors, allows determining sea water density in real time virtually, using three parameters measured in situ: temperature, hydrostatic pressure and speed of sound. Though, using equation 13 makes salinity measuring non mandatory, salinity determining can be solved using results of measuring the above three parameters to be done in several ways.

The first way is the development of special equations of (9) type for salinity. For instance, in the work published [9] this equation was developed in the following form:

$$S(T, P, c) = S_0 + \Delta S_T(T) + \Delta S_P(P) + \Delta S_c(c) + \Delta S_{TPc}(T, P, c), \quad (16)$$

where $S_0 = 0$, and $\Delta S_T$, $\Delta S_P$, $\Delta S_c$ and $\Delta S_{TPc}$ – polynomials according to independent parameters degrees: $T$, $P$ and $c$ individually and collectively, respectively. In work [9] a new set of three equation options is suggested in the form (16) with 81, with 76 and with 69 members.

Another way suggests using different, already existing or newly developed equations for the sound velocity of the following type (ref. to [4] and other):

$$c = f_c(S, T, P). \quad (17)$$

Equation of (17) type is not evident with respect to salinity, though at known values of sound velocity and pressure it can be solved with respect of salinity using known mathematical methods.

One more way can be used when sea water salinity values are determined using density values equations of the following type:

$$\rho = f_\rho(S, T, P). \quad (18)$$

Equations of this type can be found in works [4, 8]. Consequently, salinity values can be calculated with known values of density, temperature and pressure based on equation of (18) type using reverse methods. The density values used in this case, can be obtained either experimentally, or based on calculation using sea water density equation I for technology purposes, or using density equation II for research purposes which are given in this article.

Thus, the approach suggested by the authors and employing temperature, pressure and sound velocity as independent variables, opens up new possibilities in oceanographic research. Equations of (13) type can be very useful for maps generation in dynamic topography of the world ocean surface, and calculation of geostrophic currents, etc.

Equations of (13) type given in this article, together with equations of (9), (17) or (18) type allow significant simplification of procedure for salinity determining and doing without measuring of sea water relative conductivity.

Using the new equation of (13) type for sea water density is subject to the input parameters measurement uncertainty influence on calculation value accuracy. For the purpose of analyzing equation (13), we calculated the uncertainties of the results of indirect density measurements for several classes of instruments using equation II for scientific purposes. Classification of oceanographic instruments based on accuracy degree of measuring channels installed on them, is given in table 7 [16].

**Table 7.** Classification of oceanographic instruments based on accuracy degree of measuring channels installed on them [16]

| Requirements level (class) | Measurements uncertainty | | | Pressure (max) |
|---|---|---|---|---|
| | temperature | pressure | sound velocity | |
| | $u(T)$,°C | $u(P)$,$10^{-2}$ MPa**) | $u(c)$, m/s | $P_{\max}$ , MPa |
| WOCE*) | 0.002 | 3 | 0.02 | 60 |
| (1) Highest | 0.002 | 1.5 | 0.02 | 30 |
| (2) Middle | 0.005 | 1 | 0.02 | 20 |
| (3) Minimal | 0.01 | 1 | 0.02 | 10 |

*) WOCE: World Ocean Circulation Experiment; **) $10^{-2}$ MPa = 1 dbar

Calculation of uncertainty components and overall standard uncertainty of density measurements was performed using formulas:

$$u_\Sigma(\rho) = \sqrt{u_\Sigma^2(\rho)} = \sqrt{\sum_{i=1}^{3} u_i^2(\rho)} = \sqrt{\sum_{i=1}^{3} Z_i^2}\,, \qquad (17)$$

where, $u_1(\rho) = |Z_1| = u_T(\rho) = |Z_T| = 0.5|\rho( T +u(T), P, c) - \rho( T -u(T), P, c)|$ ,

$u_2(\rho) = |Z_2| = u_P(\rho) = |Z_P| = 0.5|\rho( T, P +u(P), c) - \rho( T, P -u(P), c)|$ ,

$u_3(\rho) = |Z_3| = u_c(\rho) = |Z_c| = 0.5|\rho( T, P, c +u(c)) - \rho( T, P, c - u(c))|$ .

The results of uncertainty estimates calculations of density measurements are given in Table 8.

**Table 8.** Uncertainty estimates of density measurements for different classes of instruments

| Requirements level | Temperature range | Pressure | Estimates of uncertainty components and overall standard uncertainty of density measurements, kg/m³ | | | |
|---|---|---|---|---|---|---|
| (class) | $\Delta T$, °C | $P$, MPa | $u_T(\rho)$ | $u_P(\rho)$ | $u_c(\rho)$ | $u_\Sigma(\rho(T,P,c))$ |
| WOCE | -2 - +10 | 60 | 0.007–0.009 | 0.017–0.039 | 0.010–0.016 | 0.020–0.043 |
| (1) | -2 - +18 | 30 | 0.005–0.015 | 0.007–0.012 | 0.011–0.014 | 0.014–0.019 |
| (2) | -2 - +22 | 20 | 0.011–0.015 | 0.004–0.007 | 0.012–0.014 | 0.019–0.020 |
| (3) | -2 - +26 | 10 | 0.020–0.031 | 0.004–0.007 | 0.012–0.014 | 0.026–0.033 |

From Table 8 it follows that oceanographic instruments of classes (1) and (2) have the lowest level of estimates of the overall standard uncertainty of density measurements. An additional increase in accuracy or a reduction in the overall standard uncertainty of density measurements is possible by increasing the measuring channels accuracy. We performed calculations for an increased level of requirements for the measuring channels accuracy. For example, Table 9 shows the results of uncertainty estimates calculations of density measurements for the case when the measuring channels for temperature, pressure, and sound velocity have measurement uncertainty $u(T)$ = 0.002 °C, $u(P)$ = $0.5*10^{-2}$ MPa and $u(c)$ = 0.01 m/s, respectively.

**Table 9.** Uncertainty estimates of density measurements for an increased level of requirements

| Temperature range | Pressure | Estimates of uncertainty components and overall standard uncertainty of density measurements, kg/m$^3$ | | | |
|---|---|---|---|---|---|
| ΔT, °C | P, MPa | $u_T(\rho)$ | $u_P(\rho)$ | $u_c(\rho)$ | $u_\Sigma(\rho(T,P,c))$ |
| -2 - +10 | 60 | 0.007–0.009 | 0.003–0.007 | 0.005–0.008 | 0.009–0.013 |
| -2 - +18 | 30 | 0.005–0.006 | 0.005–0.006 | 0.006–0.007 | 0.002–0.004 |
| -2 - +22 | 20 | 0.004–0.006 | 0.002–0.004 | 0.006–0.007 | 0.008–0.009 |
| -2 - +26 | 10 | 0.004–0.006 | 0.002–0.004 | 0.006–0.007 | 0.008–0.009 |

Thus, using the equation of state of a new type (13), it is also possible to develop requirements for the accuracy levels of temperature, pressure, and sound velocity measuring channels when designing relevant oceanographic instruments, depending on the requirements of the tasks for which they are intended.

Figure 4 shows an example of calculating the isolines of seawater density using to equation II for research purposes.

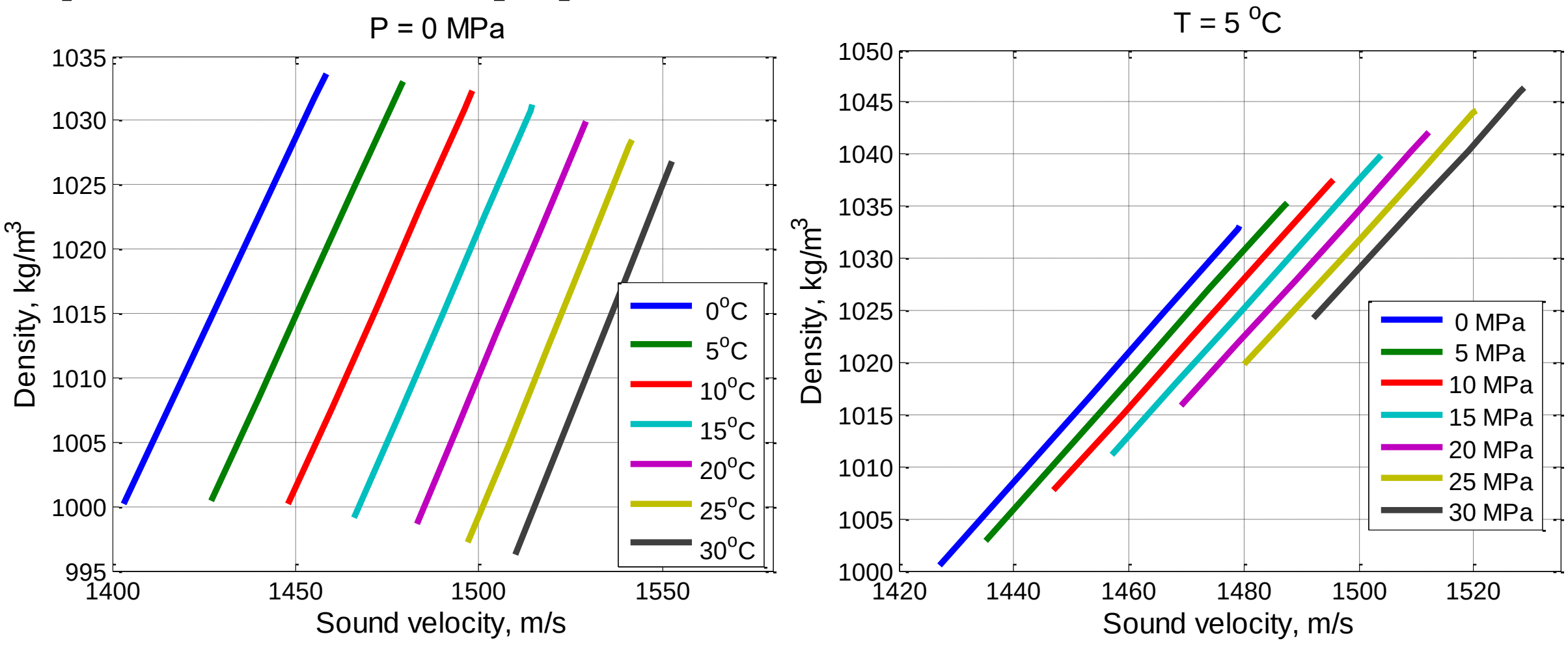

**Figure 4:** dependence of density on the sound velocity using equation II on isolines when temperature and pressure are constant

We should note that density isolines $\rho(T, P, c)$ when $T$=const and $P$=const are virtually linear dependent on the sound velocity.

## Conclusion.

The new kind of equations intended for interpolation of the sea water density depending on the measured parameters - temperature, pressure and sound velocity has been justified theoretically and implemented in practice.

For different ranges of the sea parameters two equations of the new kind for the sea water density have been developed.

Equation I is suitable for technical calculations in a wide range of measured parameters and can be used to calculate the density of anomalous, fresh and slightly saline waters at temperatures up to 40°C.

Equation II is more accurate and can be used for scientific calculations in a typical oceanographic parameter range. Both the equations can be recommended for inclusion in the software of acoustic profilographs (SVP-instruments) which must have the uncertainty of the measuring channels allowing determining the density of seawater in situ with the required accuracy